# TITLE

## Using Gamma Functions in the Mathematical Formulation of the Impact Crater Size-Age Frequency Distribution on Earth and Mars.

Author: William F Bruckman

### Abstract

A review of a mathematical formulation that describes the number of impact craters as function of diameter and time of formation is presented, where the use of Gamma functions is emphasized. The application of this formalism for the description of the impact crater data of Planets Earth and Mars is also discussed.

### 1. Introduction

When solving differential or integral equations an ideal outcome is to express the solution in terms of elementary or special functions. In that case the mathematical and physical interpretation of the solutions is clarified. Moreover, with the use of algebraic computing, the comparison of the prediction of theoretical models with the observational data is greatly facilitated.

This paper will consider work in reference1 (Earth and Mars Crater Size Frequency Distribution and Impact Rates: Theoretical and Observational Analysis; William Bruckman, Abraham Ruiz, and Elio Ramos; Arxiv: 1212.3273), which presented a theoretical formulation describing impact crater data on Earth and Mars, giving the number of craters as functions of diameter, and time of formation, successfully reproducing the observations. The revision will emphasize the presentation of the solutions of the models in terms of Gamma functions.

### 2. General Considerations

Impact craters, of a given diameter $D$, are formed at a certain rate $\Phi$, and are also depleted, as they get older, by a variety of processes, at a rate proportional to their already existing number of craters, $N$. Hence, the number of craters eliminated in the time interval $dt$ can be express as $CNdt$, where $C$ is a parameter representing the rate of elimination per crater. On the other hand, in this time interval we also have that the number of craters produced by impacts is $\Phi dt$, and thus the net change in the number of crater numbers, $dN$, is given by

$$dN = \Phi dt - CN dt = \left(\frac{\Phi}{C} - N\right) C dt. \qquad (1)$$

This equation is expected to represent well the observational data if the number of craters is large enough to justify the assumptions that analytical mathematical continuity is a good approximation to the discrete and probabilistic nature of the problem.

We see from Eq. (1) that $N=$ constant implies that

$$N = \frac{\Phi}{C} = \Phi \tau_m, \qquad (2)$$

$$\tau_m \equiv 1/C. \qquad (3)$$

In this situation (saturation) the number of craters produced by impacts is equal to the number of craters eliminated. The dimension of $\tau_m$ is time, and we will see later in this section that this time is related to the concept of "craters mean life."

Equation (1) was integrated in reference (1) to obtain

$$N(D, 0, \tau) = \int_0^\tau \{\Phi(D, \tau`) \, Exp[-\bar{C}\tau`]\} d\tau`. \qquad (4)$$

$$\bar{C} \equiv \frac{\int_0^\tau C d\tau`}{\tau}, \qquad (5)$$

where $\bar{C}$ is the time average of $C$, and $N(D, 0, \tau)$ defined in Eq. (4) denotes the number of craters of diameter $D$, per bin size, observed at the present time ($\tau` = 0$), with age younger than $\tau$. Accordingly, defining the term "per bin", we have that the integral

$$\widetilde{N}(D_i, D_f, 0, \tau) \equiv \int_{D_i}^{D_f} N(D, 0, \tau) dD, \qquad (6)$$

gives the total number of craters with diameters in the interval between $D_i$ and $D_f$, observed at the present time, with age of formation younger than $\tau$. Also, $\Phi(D, \tau)$ is the rate of meteorite impacts, per bin, forming craters of diameter $D$ at time $\tau$, so that $\Phi_C$:

$$\Phi_C(D_i, D_f, \tau) = \int_{D_i}^{D_f} \{\Phi(D, \tau)\} dD, \qquad (7)$$

Is the cumulative impact rate of formation of craters with diameters in the interval between $D_i$ and $D_f$. For instance, if $D_f \to \infty$, which is of common use, the above integral is the total cumulative impact rate of formation of craters with diameters larger than $D_i$.

Equations (6) and (4) can be generalized so that the lower $\tau$ limit is different from zero:

$$\widetilde{N}(D_i, D_f, \tau_i, \tau_f) \equiv \int_{D_i}^{D_f} N(D, \tau_i, \tau_f) dD, \qquad (8)$$

where

$$N(D, \tau_i, \tau_f) = \int_{\tau_i}^{\tau_f} \{\Phi(D, \tau`) \, Exp[-\bar{C}\tau`]\} d\tau`. \qquad (9)$$

Thus, Eqs. (8) and (9) refer to craters with ages between $\tau_i$ and $\tau_f$.

Further discussion and applications of Eq. (8) to the Earth's crater record will be continued in Section 4. For the planet Mars, however, we will be applying Eq. (4) in the next section, but now continue its interpretation below.

Since the quantity $\Phi(D, \tau`)d\tau`$. Is the number of craters formed at time $\tau`$, during the interval $d\tau`$, and the integrand in Eq. (4): $\Phi(D, \tau`)d\tau` Exp[-\bar{C}\tau`]$, is the number of these craters, of age $\tau`$, that remain at the present time, then the expression $Exp[-\bar{C}\tau`]$ represents the fraction of these formed craters that survive after the time $\tau`$. It is then usual to call the inverse of $\bar{C}$ "the mean life": $\tau_{mean}$,

$$\tau_{mean} \equiv \frac{1}{\bar{C}} \; ; \; 1/\tau_{mean} = \bar{C} \equiv \frac{\int_0^\tau C d\tau`}{\tau} = \frac{\int_0^\tau (\frac{1}{\tau_m}) d\tau`}{\tau}. \qquad (10)$$

Thus, in this context $\tau_{mean}$ can be viewed as the mean life of craters of diameter $D$. Also, this interpretation suggests thinking of $\Phi$ as a probability of impact, rather than an impact flux, thus emphasizing the statistical nature of the impacts of asteroids and comets. Conversely, if we start with the definition of $Exp[-\bar{C}\tau`]$ as the fraction of craters surviving after the interval $\tau`$ from their formation, then we can construct Eq. (4) to represent the sum of all the contributions, to the present number, for all times $\tau`$ younger than $\tau$, and then find that $N$ satisfies the differential equation implied in Eq. (1).

Consider the following definition:

$$T(D, \tau,) \equiv \int_0^\tau C d\tau` = \bar{C} \, \tau = \tau/\tau_{mean} \quad . \qquad (11)$$

Hence $T$ is a dimensionless time that measures the numbers of mean-life in an interval $\tau$. From Eq. (11) it follows that

$$dT/d\tau = C(D, \tau) \quad , \qquad (12)$$

where $D$ is considered here as a constant parameter. Since crater elimination is a decay process, where $C$ is strictly positive, we have

$$dT/d\tau > 0 \, . \qquad (13)$$

Consequently, the function $T(D, \tau,)$ can be inverted to express $\tau$ as a function of $T$ and $D$: $\tau(D, T)$. Likewise, $C$ and $\Phi$ are each expressible as functions of $T$ and $D$. We can then rewrite Eq. (4), using Eqs. (11), (12) and (3), in the form

$$N(D, 0, \tau) = \int_0^\tau \{\Phi(D, \tau`) \, Exp[-\bar{C}\tau`]\}d\tau` = \int_0^T \{(\tfrac{\Phi}{C}) \, Exp[-T`]\}dT` =$$
$$\int_0^T \{(\Phi\tau_m) \, Exp[-T`]\}dT`. \tag{14}$$

where, in the right-hand side of Eq. (14), $\tfrac{\Phi}{C} = \Phi\tau_m$ is considered now a function of $T$, and the parameter $D$. For instance, if $\Phi\tau_m$ is a sum like

$$\Phi\tau_m = \Sigma a_s T^s \, , \, a_s \text{ and } s \text{ are independent of } T, \tag{15}$$

then we have, from Eq. (14),

$$N(D, 0, T) = \Sigma a_s \int_0^T \{T`^s \, Exp[-T`]\}dT` = \Sigma a_s \, \gamma(s + 1, T) \, , \tag{16}$$

where the lower incomplete gamma function notation was used above:

$$\gamma(s + 1, T) = \int_0^T \{T`^s \, Exp[-T`]\}dT`. \tag{17}$$

If $s$ is a whole number, as in a Taylor-Maclaurin series, we can also write

$$\gamma(s + 1, T) = s! \, (1 - e^{-T} \sum_{k=0}^{s} T^k / k! \, ). \tag{18}$$

This is our first encounter with the use of gamma functions expressing the number of craters as a function of diameter and age. We will see further use of gamma functions when considering applications to Earth's impact crater data in Section (4). We will now focus our attention on applications of Eq. (14) to the planet Mars.

### 3. Applications to the Crater-Size Frequency Distribution of Mars

It was discussed in Section 2 that the product $\tfrac{\Phi}{C} = \Phi\tau_m$ represents the value of $N$ when the production and the elimination of craters are equal and $dN = 0$. Then in a steady state situation we will have $N = \Phi\tau_m = $ constant. However, in general, $\Phi\tau_m$ could depend on time, since both $\Phi$ and $\tau_m$ could depend on time. On the other hand, since $C$ is by definition the rate of crater elimination per number of craters, we have then that $\tau_m \equiv 1/C$ is strongly influenced by the elimination of old craters due to impacts forming new craters. Therefore, an increase or decrease of $\Phi$ would be correlated with an increase or decrease of $C$. Consequently, if obliterations by impacts are important, the changes in time of $\tfrac{\Phi}{C} = \Phi\tau_m$ are smoothed out relative to the individual changes in time of $\Phi$, $C$, or $\tau_m$. In such a heuristic and realistic situation, a model in which it is assumed that $\tfrac{\Phi}{C} = \Phi\tau_m$ is constant should be a good representation of the observations. In this case, Eq. (14) becomes

$$N = \Phi\tau_m(1 - e^{-T}) \tag{19}$$

With the above simple model we were able to represent (reference 1) remarkably well the pioneering Mars crater database catalog of Barlow (1988), as illustrated in Fig. (1). Also, Fig. 2 compares the model with the more recent Mars data catalog of Robbins and Hynek (2012), and also the model is in very good agreement with observations (Bruckman 2019). The values of $\Phi\tau_m$ and $T$ for Barlow's model are

$$\Phi\tau_m = \frac{1.43 \times 10^5}{D^{1.8}}, \tag{20}$$

$$T = \bar{C}\tau = \tau/\tau_{mean} = \frac{2.48 \times 10^4}{D^{2.5}}, \tag{21}$$

and then Eq. (19) becomes

$$N = \Phi\tau_m(1 - e^{-T}) = \frac{1.43 \times 10^5}{D^{1.8}}\left(1 - Exp\left[-\frac{2.48 \times 10^4}{D^{2.5}}\right]\right), \tag{22}$$

where the unit of $D$ is kilometers. It can also be shown (Appendix A), using the assumption that $\Phi\tau_m$ is independent of $T$, that

$$\bar{\Phi}\tau = \int_0^\tau \Phi d\tau = \Phi\tau_m T = \frac{3.55 \times 10^9}{D^{4.3}}, \tag{23}$$

where $\bar{\Phi}$ is the time average of $\Phi$, and $\bar{\Phi}\tau$ is the total number of craters, of diameter $D$, per bin, created over the total time of production $\tau$. The corresponding expression for the number of craters created with diameters in the interval between $D_i$ and $D_f$ is then:

$$\tau\bar{\Phi}_C(D_i, D_f, \tau) = \int_{D_i}^{D_f} \bar{\Phi}\tau \, dD = (3.55/3.3)10^9 \left(\frac{1}{D_i^{3.3}} - \frac{1}{D_f^{3.3}}\right). \tag{24}$$

It is common to take the upper limit $D_f$ to be infinite to obtain the total number of craters produced larger than $D_i$:

$$\tau\bar{\Phi}_C(D_i, \infty, \tau) = 1.076 \times 10^9 \left(\frac{1}{D_i^{3.3}}\right), \tag{25}$$

or

$$\bar{\Phi}_C(D_i, \infty, \tau) = (1.076 \times 10^9/\tau)\left(\frac{1}{D_i^{3.3}}\right), \tag{26}$$

where $\bar{\Phi}_C(D_i, \infty, \tau)$ is the time average of the cumulative impact rate for the formation of craters larger than $D_i$. For instance, it is interesting to note that for $D_i = 1$ km, approximately $10^9$ such impacts were produced. Therefore, assuming that the total time of crater production $\tau$ was 3000 to 4000 million years, we get an average of one impact, making craters larger than 1 km, approximately every three to four years. Since the energy associated to impacts with a diameter of 1 km is close to one megaton, this result is of concern for explorations of Mars, assuming that the present impact flux average is comparable to that given by Eq. (26).

It is expected that also the corresponding impact rate for Earth has, similar to Mars, a crater diameter dependency of the form $\frac{1}{D_i^{3.3}}$, and indeed, we found that for our planet such a relation is consistent with the observations (Appendix B).

Let us continue our analysis of the implications of the above model, by looking at Eq. (21), rewritten in the form

$$\frac{\tau_{mean}}{\tau} = D^{2.5}/2.48x10^4 . \qquad (27)$$

An interesting interpretation of the above equation (Reference 1) is that it represents a proportionality relation between the mean life, of a crater of diameter $D$, and the initial volume of this crater. This conclusion is based on observations on Mars that established that the initial depths of pristine craters are proportional to $D^{k/2}$, with $k \approx 1$, and, consequently, the expected initial volumes for these craters are proportional to $D^2 D^{k/2} \approx D^{2.5}$. For instance, Garvin (2002) gives $k \approx 0.98$, while Boyce et al. (2007) give $k \approx 1.04$. Furthermore, from the application to Earth of the above formalism, to be discussed in the next section, it was concluded that craters in our planet also have their mean-life proportional to $\approx D^{2.5}$. Thus, we have that the relation $\tau_{mean}$ proportional to the crater initial volume is not only intuitively appealing, but also helps us understand why we have similar $D$ exponents in the $\tau_{mean}$ for Earth and Mars, notwithstanding these planets contrasting geological evolutions.

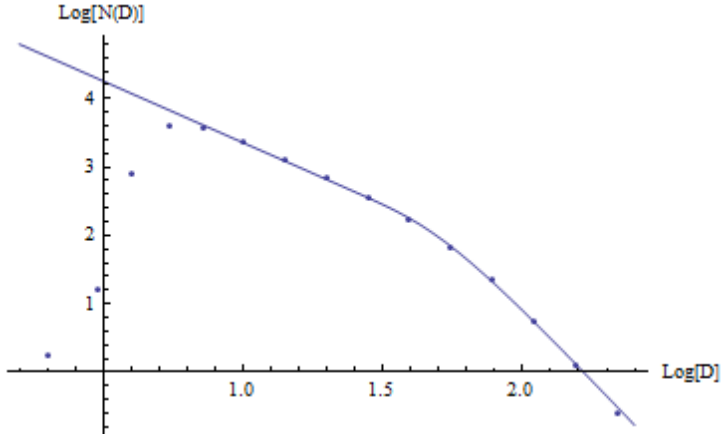

FIGURE (1): Log-Log plot of number of craters per bin, $N(D)$ vs $D$ based on Barlow's Mars catalog (1988). The number $N(D)$ is calculated by counting the number of craters in a bin $\Delta D = D_R - D_L$, and then dividing this number by the bin size. The point is placed at the mathematical average of $D$ in the bin: $(D_R + D_L)/2$. The bin size is $\Delta D = (\sqrt{2} - 1)D_L$, so that $\frac{D_R}{D_L} = \sqrt{2}$. ). The curve is from the model implied by Eq. (22). We see that the theoretical curve shown differs significantly from the observed data for $D$ less than about $8 km$. However, according to Barlow, the empirical data undercounts the actual crater population for $D$ less than $8 km$. However, more recent Mars crater data by Robbins et al. (2012) was used to update the observations, yielding similar results to the model in Figure 1, but extending the range to craters with diameters down to 1 km (see Fig. 2).

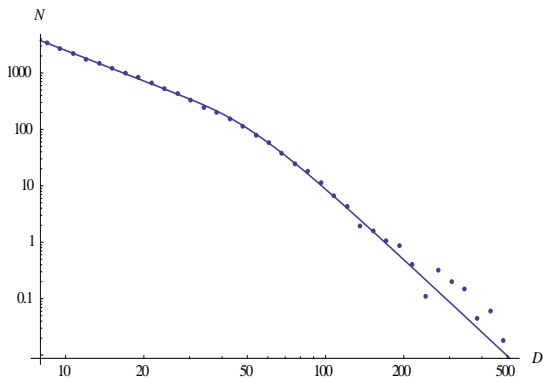

FIGURE (2): Log-Log plot of $N(D)$, vs $D$(km), based on the Mars catalog of Robbins et al (2012), (Bruckman (2019)). Bin size is $\Delta D = (\ 2^{1/6} - 1)D_L$. Note that for $D > \sim 300\ km$, the data points are above the curve of the analytic model. However, we expect that the analytical model will be less reliable when the number of craters in a given bin is so small that statistical continuous models break down. Moreover, another source of discrepancy could be that these very large craters were being formed at high proportions at older times $\tau$, thus perhaps belonging to the so-called heavy bombardment era, characterized by a much higher impact flux.

## 4. Applications to Planet Earth

The number of identified impact craters on Earth is close to 190 (Planetary and Space Science Center: PASSC.com), while, in contrast, the number of craters used for Mars in Fig. (1) was 42,284. Therefore, in the analysis of Earth's crater data it is convenient to use the cumulative number of impacts of craters, $\widetilde{N}(D_i, D_f, 0, \tau)$, defined in Eq. (6), instead of $N(D, 0, \tau)$, defined in Eq. (4). Furthermore, for $N(D, 0, \tau)$, the simplified expression in Eq. (19) will be used, since it reproduced the Martian impact data very well. Thus we have

$$\widetilde{N}(D_i, D_f, 0, \tau) \equiv \int_{D_i}^{D_f} N(D, 0, \tau) dD = \int_{D_i}^{D_f} \Phi \tau_m (1 - e^{-T}) \, dD =$$

$$\int_{D_i}^{D_f} \Phi \tau_m \, dD - \int_{D_i}^{D_f} \Phi \tau_m e^{-T} dD. \tag{28}$$

In addition, let us assume that

$$\Phi \tau_m = \frac{H}{D^{m`}}, \tag{29}$$

$$T = \bar{C}\tau = \frac{B\tau}{D^p}, \tag{30}$$

$$\bar{\Phi}\tau = \int_0^\tau \Phi \, d\tau` = \Phi \tau_m T = \frac{A\tau}{D^m}. \tag{31}$$

where $H, m`, B, p, A$ and $m$ are independent of $D$, and, from Eqs. (29), (30) and (31),

$$m = m` + p, \tag{32}$$

$$A = HB. \tag{33}$$

Equations (29), (30), and (31) are a generalization for Earth of the corresponding equations, (20), (21) and (23), describing the crater distribution for Mars. For Mars, we have $H = 1.43 \times 10^5$, $B\tau = 2.48 \times 10^4$, and $A\tau = 3.55 \times 10^9$. However, for our planet these values will have to be redetermined. Also, the exponents $m$ and $p$ should come out from the fitting to Earth data. As was discussed in previous section, a value of $m = 4.3$, in the exponent of $D$ of the impact flux $\bar{\Phi}$ is also consistent with the Earth observational impact rate data (Appendix B). The value $p = 2.5$ is also consistent with the Earth observations, to be discussed in this section.

After the substitution of the expressions in Eqs. (29), (30), and (31) in Eq. (28) the first integral in the right-hand side is elementary, hence, we will turn our attention to the second integral:

$$-\int_{D_i}^{D_f} \Phi \tau_m e^{-T} dD = -\int_{D_i}^{D_f} \frac{H}{D^{m`}} \left\{ Exp \left[ \frac{-B\tau}{D^p} \right] \right\} dD. \tag{34}$$

To emphasize that the variable of integration is now $D$, while $\tau$ is a fixed parameter, we rename $T$ as $U$:

$$T = U = \frac{B\tau}{D^p}, \tag{35}$$

or

$$D = \left[\frac{B\tau}{U}\right]^{1/p}, \tag{36}$$

from which, differentiating with respect to $D$, holding $\tau$ fixed,

$$dD = -[B\tau]^{\frac{1}{p}}[U]^{\frac{-1}{p}-1} dU/p. \tag{37}$$

Substituting Eqs. (35), (36), and (37) in Eq. (34) we get

$$-\int_{D_i}^{D_f} \Phi \tau_m e^{-T} dD = \{H/(p[B\tau]^n)\} \int_{U_i}^{U_f} U^{n-1}\{Exp[-U]\} dU = \{H/(p[B\tau]^n)\} \Gamma[n, U_i, U_f], \tag{38}$$

where

$$n \equiv (m` - 1)/p, \tag{39}$$

$$U_i = \frac{B\tau}{D_i^p}, \tag{40}$$

$$U_f = \frac{B\tau}{D_f^p}, \tag{41}$$

and

$$\Gamma[n, U_i, U_f] = \int_{U_i}^{U_f} U^{n-1}\{Exp[-U]\} dU \tag{42}$$

is the generalized incomplete gamma function. Consequently, we can rewrite Eq. (28) in the form

$$\widetilde{N}(D_i, D_f, 0, \tau) \equiv \int_{D_i}^{D_f} \frac{H}{D^m} dD + \{H/(p[B\tau]^n)\} \Gamma[n, U_i, U_f] \tag{43}$$

The above integral represents the number of craters with diameters in the interval between $D_i$ and $D_f$, that are younger than $\tau$. Hence, the number of craters formed with ages between $\tau_i$ and $\tau_f$ is

$$\widetilde{N}(D_i, D_f, \tau_i, \tau_f) \equiv \int_{D_i}^{D_f} N(D, \tau_i, \tau_f) dD = \widetilde{N}(D_i, D_f, 0, \tau_f) - \widetilde{N}(D_i, D_f, 0, \tau_i) =$$

$$\{H/(p[B\tau_f]^n)\} \Gamma\left[n, \frac{B\tau_f}{D_i^p}, \frac{B\tau_f}{D_f^p}\right] - \{H/(p[B\tau_i]^n)\} \Gamma\left[n, \frac{B\tau_i}{D_i^p}, \frac{B\tau_i}{D_f^p}\right]. \tag{44}$$

Here, if $B$ is a function of time, it should be evaluated at the corresponding $\tau_i$ or $\tau_f$.

Another useful concept is the statistical mean of a function of $D: f(D)$, which is defined using $N(D, \tau_i, \tau_f)$, as follows

$$\bar{f} = \int_{D_i}^{D_f} fN(D,\tau_i,\tau_f)\, dD / \{\int_{D_i}^{D_f} N(D,\tau_i,\tau_f)\, dD\} = \int_{D_i}^{D_f} fN(D,\tau_i,\tau_f)\, dD / \{\widetilde{N}(D_i,D_f,\tau_i,\tau_f)\}. \quad (45)$$

For instance, if $f = D$ we get, from definition (45), the average diameters of craters with diameters and ages in the intervals $D_i \leq D \leq D_f$, and $\tau_i \leq \tau \leq \tau_f$, respectively. In this case it follows that the numerator of Eq. (45) is

$$\int_{D_i}^{D_f} DN(D,\tau_i,\tau_f)\, dD = \{H/(p[B\tau_f]^{n`})\}\, \Gamma\left[n`, \frac{B\tau_f}{D_i^p}, \frac{B\tau_f}{D_f^p}\right] - \{H(p[B\tau_i]^{n`})\}\Gamma\left[n`, \frac{B\tau_i}{D_i^p}, \frac{B\tau_i}{D_f^p}\right], \quad (46)$$

where

$$n` \equiv \frac{m`-2}{p} = n - 1/p \ . \quad (47)$$

Hence

$$\bar{D} = [\frac{1}{\widetilde{N}(D_i,D_f,\tau_i,\tau_f)}][\{H/(p[B\tau_f]^{n`})\}\, \Gamma\left[n`, \frac{B\tau_f}{D_i^p}, \frac{B\tau_f}{D_f^p}\right] - \{H/(p[B\tau_i]^{n`})\}\, \Gamma\left[n`, \frac{B\tau_i}{D_i^p}, \frac{B\tau_i}{D_f^p}\right]] \quad (48)$$

The above expression was adapted and applied to the Earth crater data in reference (1). The value of $p$ was determined by the best fitting of the data to the model given in Eq. (48), and yielded a value similar to that for Mars. As stated, this is interpreted to be the result of the proportionality of $\tau_{mean}$ to the initial volume of craters, and that this volume is in turn proportional to $D^p$. From this fitting to observation also came an approximate value for Earth's parameter $B$.

The expression $\widetilde{N}(D_i, D_f, \tau_i, \tau_f)$ in Eq. (44), was also used in reference (1) to describe the number of Earth's craters as a function of diameter and age, as illustrated in figures C1 and C2 in Appendix C. The values $p = 2.5$, $m = 4.3$, and $m` = m - p = 1.8$ were assumed since they were observationally justified. The value of $H = A/B$ was also needed, and, since $B$ was estimated from observations of $\bar{D}$, then the value of $A$ remained to be estimated, as described in Appendix C. A remarkable agreement of the model with observations was obtained.

## Appendix A

The number of impacts, during the time $\tau$, producing craters of diameter $D$, per bin, can be expressed as

$$\bar{\Phi}\tau = \int_0^\tau \Phi d\tau` = \int_0^\tau \Phi\tau_m d\tau`/\tau_m .\qquad\text{A1}$$

Using Eqs. (3) and (12), we get

$$dT/d\tau = C(D,\tau) = \frac{1}{\tau_m}.\qquad\text{A2}$$

We can then rewrite the right hand side of Eq. (A1) in the form

$$\int_0^\tau \Phi\tau_m d\tau`/\tau_m = \int_0^T \Phi\tau_m dT` .\qquad\text{A3}$$

If furthermore $\Phi\tau_m$ is independent of $T$ we have

$$\int_0^\tau \Phi\tau_m d\tau`/\tau_m = \int_0^T \Phi\tau_m dT` = \Phi\tau_m \int_0^T dT` = \Phi\tau_m T .\qquad\text{A4}$$

Therefore, from Eqs. (A1) and (A4),

$$\bar{\Phi}\tau = \int_0^\tau \Phi d\tau` = \int_0^\tau \Phi\tau_m d\tau`/\tau_m = \Phi\tau_m T.\qquad\text{A5}$$

Note also that, since

$$T = \bar{C}\tau = \tau/\tau_{mean} ,\qquad\text{A6}$$

from (A5) we get

$$\bar{\Phi}\tau_{mean} = \Phi\tau_m.\qquad\text{A7}$$

**Appendix** B

Let us investigate the observational implications of the assumption of an average impact flux for Earth given by

$$\bar{\Phi}\tau = \int_0^\tau \Phi d\tau` = \frac{A\tau}{D^{4.3}},$$  B1

which implies the following cumulative impact flux

$$\bar{\Phi}_C(D_i, \infty, \tau) = \int_{D_i}^\infty \bar{\Phi} \, dD = \frac{A/3.3}{D^{3.3}},$$  B2

where we drop the $i$ sub index from $D$, in the right-hand side of Eq. (B2). The value of $A$ can be estimated for Earth from the result of Grieve and Shoemaker (1994) for $D = 20km$:

$$\bar{\Phi}_C(20km, \infty, \tau) = \frac{(5.5 \mp 2.7)10^{-9}}{(my)km^2} 4\pi R^2 \approx 2.8[\frac{1 \mp 0.50}{my}],$$  B3

where $R$ is the Earth's radius, and $my$ is million years. Comparing Eq. (B2), evaluated at $D = 20km$, with Eq. (B3) we obtain

$$A = 9.24[1 \mp 0.50]\frac{(20)^{3.3}}{my},$$  B4

and thus

$$\bar{\Phi}_C(D_i, \infty, \tau) \equiv \bar{\Phi}_C = 2.8[\frac{1 \mp 0.50}{my}](20/D)^{3.3}$$  B5

This equation is a generalization of the result of Grieve and Shoemaker (1994), which gives the Earth's impact rate for the formation of craters with diameters larger than $D$. It incorporates the 3.3 exponent on $D$ that we deduced from the model and observations from Mars.

The diameter of a crater corresponds to an energy, $E$, associated to the impact, and hence Eq. (B5) can be re-expressed as (reference 1)

$$\bar{\Phi}_C(E) = \frac{[1 \mp 0.5]}{14.5y \, E^{0.86}},$$  B6

where $E$ is in megatons. Equation (B6) gives similar predictions to those of Poveda et al. (1999). The predictions of Eq. (B6) are also in agreement with Silber et al. (2009), that, for impacts with energies larger than a megaton, gives one Earth impact about every 15 years. It is interesting to note that, according to Eq. (B6), events like the 2013 Chelyabinsk meteorite of energy of about 0.5 megatons are predicted to happen with a periodicity near one every $8/(1 \mp 0.5)$ years, so that this type of event is expected to be repeated in the near future.

    Observations in the last few decades of lunar meteorites, called Lunar Flashes, provide a direct determination of the impact rate, at these low range of energies (see for example Oberst et al. (2012), and Suggs et al. (2014)). For instance, Oberst et al. (2012) interpreted data of lunar flashes, and concluded a rate of $10^{-3}$ impacts per $km^2$ per year, for energies $\geq \sim 8x10^{-6}$ kilotons. This result, translated to the total Earth`s surface area, becomes approximately $5.1x10^5$ impacts per year for these energies, while from Eq. (B6) we get about $6.3[1 \mp 0.5]x10^5$ impacts per year, which is consistent with the above result for lunar flashes.

**Appendix C**

To reduce the uncertainties due to undercounting in the Earth crater data we selected the following regions for the study in reference 1:

(a) Continental United States
(b) Canada up to the Arctic Circle
(c) Europe
(d) Australia

The crater data is taken from The Planetary and Space Science Centre (www.passc.net). Then, in Eq. (44), instead of using for the total Earth's impact flux

$$\overline{\Phi} = (1/\tau) \int_0^\tau \Phi d\tau` = \frac{A}{D^{4.3}}, \qquad \text{C1}$$

we used for our study the more accurate impact flux corresponding to the area under consideration above in a,b,c,d, which is given by

$$\overline{\Phi}_{acc} = \frac{A_{acc}}{D^{4.3}}, \qquad \text{C2}$$

where

$$A_{acc} \equiv A \frac{Area\ Under\ Consideration}{Earth`s\ Surface\ Area}, \qquad \text{C3}$$

where $A$ is given, from Eq. (B4), by

$$A = 9.24[1 \mp 0.50]\frac{(20)^{3.3}}{my} = (1.82)10^5[1 \mp 0.5]/my. \qquad \text{C4}$$

Accordingly, $H = A/B$ becomes $H_{acc} = \frac{A_{acc}}{B}$, with $B$ estimated from the curve $\overline{D}$ vs. crater age, given by Eq. (48), fitting to the Earth's data. Therefore, we can write the theoretical $\widetilde{N}$ with no free parameters, and compare it with the observations, as described below. We do this first in table (I) and Figure (C1), for craters with $D \geq 20km$ and cumulative age starting with $\tau = 1my$ up to $\tau = 2,000my$. Furthermore, we put $\tau_f = 2,500my$ and $D_f = 300km$, since all craters in the field of study are within this bin size. This theoretical curve, $\widetilde{N}(\tau)$, is then compared with the corresponding observational data, and the very good agreement between theory and observation is noteworthy. On the other hand, we also compare theory and observation in Table II and Figure (C2), where now $\widetilde{N}$ cumulative represents the number of craters of all ages, $1my \leq \tau \leq 2,500my$, with diameters greater than or equal to $D$. Again, the theoretical $\widetilde{N}(D)$ is in very good agreement with the observations for $D \geq \sim 20km$, although not so good for $D \leq \sim 20km$, which is as expected due to the undercounting of craters of these sizes.

**Table I**

| $\tau(my)$ | $\tilde{N}[\tau, D \geq 20km\ ]$ | Observation |
|---|---|---|
| 1 | 33.14 | 33 |
| 10 | 32.00 | 32 |
| 20 | 30.80 | 31 |
| 40 | 28.62 | 29 |
| 50 | 27.62 | 28 |
| 100 | 23.40 | 24 |
| 150 | 20.24 | 20 |
| 200 | 17.80 | 17 |
| 300 | 14.20 | 13 |
| 400 | 11.70 | 10 |
| 600 | 8.50 | 8 |
| 800 | 6.50 | 5 |
| 1000 | 5.00 | 5 |
| 1200 | 3.89 | 4 |
| 1400 | 2.99 | 3 |
| 1600 | 2.25 | 3 |
| 1800 | 1.62 | 2 |
| 2000 | 1.08 | 1 |

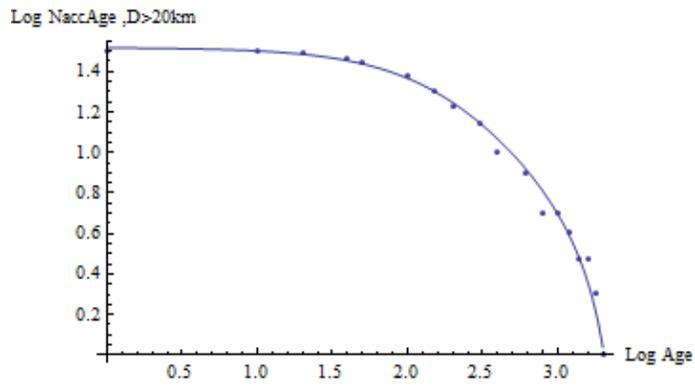

FIGURE (C1): $Log[\widetilde{N}]$ vs $Log[\tau \equiv Age]$, for all diameters $D \geq 20km$. See Table I.

**Table II**

| D | $\widetilde{N}[D, 1my \leq \tau \leq 2,500my\ ]$ | Observation |
|---|---|---|
| 1 | 166.00 | 121 |
| 2 | 165.00 | 118 |
| 4 | 137.00 | 99 |
| 8 | 82.60 | 72 |
| 16 | 42.40 | 37 |
| 20 | 33.14 | 33 |
| 32 | 18.18 | 16 |
| 45 | 10.37 | 10 |
| 64 | 4.79 | 5 |
| 91 | 1.82 | 2 |
| 128 | 0.62 | 1 |

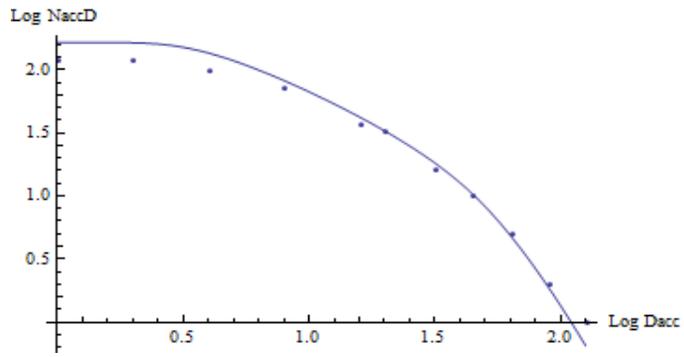

FIGURE (C2): $[Log[\tilde{N}]$ vs. $Log[D_{Acc} \equiv D]$, for all ages between $1my \leq \tau \leq 2{,}500my$ (Table II).

**References**

**1.** Bruckman, W.F., Ruiz, A., Ramos, E. (2012). Earth and Mars Crater Size Frequency Distribution and Impact Rates: Theoretical and Observational Analysis; arXiv:1212.3273(astro-ph)

2. Barlow, N.G. (1988). Icarus 75, 285.

3. Robbins, S.J., and Hynex, B.M. (2012). Global Database of Mars Impact Craters $\geq 1 km$.; Journal of Geophysical Research: Planets 117(E5)

4. Bruckman, W.F. (2019). Researchgate preprint. DOI: 10.13140/R.G.2.2.33363.43047

5. Garvin, J.B. (2002). Lunar and Planetary Science 33, 1255.

6. Boyce, J.M., Garbeil, H. (2007). Geophysical Research Letters 34(16).

7. Planetary and Space Science Centre (PASSC), Earth Impact Database (http://www.passc.net/EarthImpactDatabase/

8. Grieve and Shoemaker (1994). The Record of Past Impacts on Earth. In: *Hazards Due To Comets And Asteroids,* T. Gehrels, ed., The University of Arizona Press.

9. Poveda, A., Herrera, M.A., Garcia, J.L., Curioca, K. (1999) Planetary and Space Science 47, 679.

10. Silber, E.A., Revelle, D.O., Brown, P.G., Edwards, W.N. (2009). Journal of Geophysical Research 114, E08006.

11. Oberst, J., A., Christou, A., Suggs, R., Moser, D., Daubar, I.J., McEwenf, A.S., Burchell, M., Kawamura, T., Hiesinger, H., Wünnemann, K., Wagner, R., Robinson, M.S. (2012); The Present day Flux of Large Meteoroids on the Lunar Surface. A synthesis of Models and Observational Techniques. Planetary and Space Science 74, 179–193

12. Suggs, R.M., Muser, D.E., Cooke, W.J., Suggs, R.J. (2014). The Flux of Kilogram-Sized Meteoroids From Lunar Impact Monitoring. Icarus April 2014.